\documentstyle[12pt]{article}
\newlength{\extraspace}
\newlength{\extraspaces}
\newcommand{\be}{\begin{equation}
\addtolength{\abovedisplayskip}{\extraspaces}
\addtolength{\belowdisplayskip}{\extraspaces}
\addtolength{\abovedisplayshortskip}{\extraspace}
\addtolength{\belowdisplayshortskip}{\extraspace}}
\newcommand{\ee}{\end{equation}}
\newcommand{\ba}{\begin{eqnarray}
\addtolength{\abovedisplayskip}{\extraspaces}
\addtolength{\belowdisplayskip}{\extraspaces}
\addtolength{\abovedisplayshortskip}{\extraspace}
\addtolength{\belowdisplayshortskip}{\extraspace}}
\newcommand{\ea}{\end{eqnarray}}
\newcommand{\nonu}{\nonumber \\[.5mm]}
\newcommand{\A}{&\!\!\!}
\begin{document}
\thispagestyle{empty}
\setlength{\baselineskip}{6mm}
%
%
\vspace*{7mm}
\begin{center}
{\large\bf On the linearization of nonlinear supersymmetry \\[2mm]
based on the commutator algebra 
} \\[20mm]
%
%
{\sc Motomu Tsuda}
\footnote{
\tt e-mail: tsuda@sit.ac.jp} 
\\[5mm]
{\it Laboratory of Physics, 
Saitama Institute of Technology \\
Fukaya, Saitama 369-0293, Japan} \\[20mm]
\begin{abstract}
We discuss a linearization procedure of nonlinear supersymmetry (NLSUSY) 
based on the closure of the commutator algebra for variations of functionals of Nambu-Goldstone fermions 
and their derivative terms under NLSUSY transformations in Volkov-Akulov NLSUSY theory. 
In the case of a set of bosonic and fermionic functionals, which leads to (massless) vector linear supermultiplets, 
we explicitly show that general linear SUSY transformations of basic components defined from those functionals 
are uniquely determined by examining the commutation relation in the NLSUSY theory. 
\\[5mm]
%
%
\end{abstract}
\end{center}

\newpage

\noindent
The Volkov-Akulov (VA) nonlinear supersymmetric (NLSUSY) theory \cite{VA} describes 
dynamics of (massless) Nambu-Goldstone (NG) fermions which are inevitably produced from a spontaneous SUSY breaking. 
It also relates to linear SUSY (LSUSY) theories and the relations between the VA NLSUSY theory and LSUSY ones 
are explicitly shown, e.g. for $N = 1$ and $N = 2$ SUSY \cite{IK}-\cite{STT2}. 
In the relations (NL/LSUSY relations), component fields in LSUSY theories are expressed 
as functionals (composites) of the NG fermions and their derivative terms, 
which lead to LSUSY transformations of the component fields under NLSUSY ones of the NG fermions. 
The fundamental action in the VA NLSUSY theory and LSUSY actions (with $D$ terms) 
for linear supermultiplets are related to each other through the functionals of the NG fermions. 

On the other hand, the NLSUSY general relativistic (GR) theory has been constructed \cite{KS,ST1} 
as a generalization of the NLSUSY theory to curved spacetime, 
where a NLSUSY Einstein-Hilbert type action is defined in terms of a vierbein and the NG fermions. 
The fundamental action in the NLSUSY-GR theory has symmetries which are isomorphic to $SO(N)$ super-Poincar\'e group 
and it contains the VA NLSUSY action in the cosmological term, 
where a dimensional constant in the NLSUSY theory is expressed by using the cosmological and gravitational constants. 
Therefore, the NL/LSUSY relations in flat spacetime contribute to 
the understanding of the low-energy physics in the NLSUSY GR theory 
and it is an interesting and important problem to know more explicitly the NL/LSUSY relations for $N \ge 2$ SUSY 
including up to general auxiliary fields in linear supermultiplets. 

In order to investigate the NL/LSUSY relations in more detail, 
it is a useful method to identify basic components for general linear supermultiplets 
by considering variations of functionals of the NG fermions based on a commutator algebra for the NLSUSY transformations 
in the VA NLSUSY theory and by constructing LSUSY transformations of bosonic and fermionic components, 
which satisfy the same commutation relation. 
In this letter, we discuss the linearization procedure of NLSUSY in extended SUSY 
based on the fact in flat spacetime that every functional of the NG fermions and their derivative terms 
satisfies the commutator algebra in the VA NLSUSY theory under the NLSUSY transformations. 
By introducing a set of bosonic and fermionic functionals, which leads to (massless) vector supermultiplets, 
we show that general LSUSY transformations of basic components defined from those functionals 
are uniquely determined by examining the commutation relation in the NLSUSY theory. 

The fundamental action in the VA NLSUSY theory \cite{VA} is defined in terms of (Majorana) NG fermions $\psi^i$ as 
\be
S_{\rm NLSUSY} = - {1 \over {2 \kappa^2}} \int d^4x \ \vert w \vert, 
\label{NLSUSYaction}
\ee
where $\kappa$ is a dimensional constant whose dimension is (mass)$^{-2}$ 
and the determinant $\vert w \vert$ is 
\footnote{
The indices $i,j, \cdots = 1, 2, \cdots, N$ 
and Minkowski spacetime indices are denoted by $a, b, \cdots = 0, 1, 2, 3$. 
Gamma matrices satisfy $\{ \gamma^a, \gamma^b \} = 2 \eta^{ab}$ 
with the Minkowski spacetime metric $\eta^{ab} = {\rm diag}(+,-,-,-)$ 
and $\displaystyle{\sigma^{ab} = {i \over 4}[\gamma^a, \gamma^b]}$ is defined. 
}
\be
\vert w \vert = \det(w^a{}_b) = \det(\delta^a_b + t^a{}_b) 
\ee
with $t^a{}_b = - i \kappa^2 \bar\psi^i \gamma^a \partial_b \psi^i$. 
The NLSUSY action (\ref{NLSUSYaction}) is invariant under the NLSUSY transformations of $\psi^i$, 
\be
\delta_\zeta \psi^i = {1 \over \kappa} \zeta^i + \xi^a \partial_a \psi^i, 
\label{NLSUSY}
\ee
where $\xi^a = i \kappa \bar\psi^j \gamma^a \zeta^j$ with constant (Majorana) spinor parameters $\zeta^i$. 
The commutator algebra for the NLSUSY transformations (\ref{NLSUSY}) closes as 
\be
[\delta_{\zeta_1}, \delta_{\zeta_2}] = \delta_P(\Xi^a), 
\label{NLSUSYcomm}
\ee
where $\delta_P(\Xi^a)$ means a translation with parameters $\Xi^a = 2 i \bar\zeta_1^i \gamma^a \zeta_2^i$. 

It is also shown that every Lorentz-tensor (or scalar) functionals of $\psi^i$ and their derivative terms 
satisfies the commutator algebra (\ref{NLSUSYcomm}) under the NLSUSY transformations (\ref{NLSUSY}): 
If we define bosonic or fermionic functionals of $\psi^i$ 
and their derivative terms ($\partial \psi^i$, $\partial^2 \psi^i$, $\cdots$, $\partial^n \psi^i$), 
in which $\gamma$-matrices are also used, as 
\be
F^I_A = F^I_A(\psi^i, \partial \psi^i, \partial^2 \psi^i, \cdots, \partial^n \psi^i) 
\label{functionals}
\ee
with $A$ meaning the Lorentz indices of $(a, ab, \cdots, {\rm etc.})$ 
and $I$ being the (internal) ones of $(i, ij, \cdots, {\rm etc.})$, 
then the commutator algebra for variations of the functionals $F^I_A$ (or $F_A$, $F^I$ and $F$) 
under the NLSUSY transformations (\ref{NLSUSY}) closes as 
\be
[\delta_{\zeta_1}, \delta_{\zeta_2}] F^I_A = \Xi^a \partial_a F^I_A. 
\label{NLSUSYcomm2}
\ee
This relation (\ref{NLSUSYcomm2}) is proved from the fact that 
the derivative terms ($\partial \psi^i$, $\partial^2 \psi^i$, $\cdots$, $\partial^n \psi^i$) 
and products of two kinds of the fuctionals $F^I_A$ and $G^J_B$ which are respectively defined 
as Eqs.(\ref{functionals}) and (\ref{NLSUSYcomm2}) satisfy the same commutation relation 
(for example, see \cite{ST2}). 

Here let us introduce bosonic and fermionic functionals which are expressed 
as the following products of $(\psi^i)^{2(n-1)}$-terms and $\vert w \vert$ 
or $(\psi^i)^{2n-1}$-ones and $\vert w \vert$  ($n = 1, 2, \cdots$); namely, we define 
\be
b^i{}_A{}^{jk}{}_B{}^{l \cdots m}{}_C{}^n \left( (\psi^i)^{2(n-1)} \vert w \vert \right) 
= \kappa^{2n-3} \bar\psi^i \gamma_A \psi^j \bar\psi^k \gamma_B \psi^l \cdots \bar\psi^m \gamma_C \psi^n \vert w \vert, 
\label{bosonic}
\ee
meaning $b = \kappa^{-1} \vert w \vert$, \ $b^i{}_A{}^j = \kappa \bar\psi^i \gamma_A \psi^j \vert w \vert$, 
\ $b^i{}_A{}^{jk}{}_B{}^l = \kappa^3 \bar\psi^i \gamma_A \psi^j \bar\psi^k \gamma_B \psi^l \vert w \vert$, 
$\cdots$ etc., and 
\be
f^{ij}{}_A{}^{kl}{}_B{}^{m \cdots n}{}_C{}^p \left( (\psi^i)^{2n-1} \vert w \vert \right) 
= \kappa^{2(n-1)} \psi^i \bar\psi^j \gamma_A \psi^k \bar\psi^l \gamma_B \psi^m \cdots \bar\psi^n \gamma_C \psi^p \vert w \vert, 
\label{fermionic}
\ee
meaning $f^i = \psi^i \vert w \vert$, \ $f^{ij}{}_A{}^k = \kappa^2 \psi^i \bar\psi^j \gamma_A \psi^k \vert w \vert$, 
$\cdots$ etc. 
In these functionals, (Lorentz) indices $A, B, \cdots$ are used as ones for a basis of $\gamma$ matrices, 
i.e. $\gamma_A = {\bf 1}, i \gamma_5, i \gamma_a, \gamma_5 \gamma_a$ or $\sqrt{2} i \sigma_{ab}$. 
We note that $f^i$ give the leading order of superchages $Q^i$ 
and the definitions of the functionals (\ref{bosonic}) and (\ref{fermionic}) terminate 
with $n = 2N + 1$ and $n = 2N$, respectively, because $(\psi^i)^{4N+1} = 0$. 

Then, the variations of the functionals (\ref{bosonic}) and (\ref{fermionic}) 
under the NLSUSY transformations (\ref{NLSUSY}) become 
\ba
\delta_\zeta b^i{}_A{}^{jk}{}_B{}^{l \cdots m}{}_C{}^n 
\A = \A 
\kappa^{2(n-1)} \left[ \left\{ \left( \bar\zeta^i \gamma_A \psi^j + \bar\psi^i \gamma_A \zeta^j \right) 
\bar\psi^k \gamma_B \psi^l \cdots \bar\psi^m \gamma_C \psi^n + \cdots \right\} \vert w \vert \right. 
\nonu
\A \A 
\left. + \kappa \partial_a \left( \xi^a \bar\psi^i \gamma_A \psi^j \bar\psi^k \gamma_B \psi^l 
\cdots \bar\psi^m \gamma_C \psi^n \vert w \vert \right) \right], 
\label{variation1}
\\[1mm]
\delta_\zeta f^{ij}{}_A{}^{kl}{}_B{}^{m \cdots n}{}_C{}^p
\A = \A 
\kappa^{2n-1} \left[ \left\{ \zeta^i \bar\psi^j \gamma_A \psi^k \bar\psi^l \gamma_B \psi^m \cdots \bar\psi^n \gamma_C \psi^p 
\right. \right. 
\nonu
\A \A 
\left. + \psi^i \left( \bar\zeta^j \gamma_A \psi^k + \bar\psi^j \gamma_A \zeta^k \right) 
\bar\psi^l \gamma_B \psi^m \cdots \bar\psi^n \gamma_C \psi^p + \cdots \right\} \vert w \vert 
\nonu
\A \A 
\left. + \kappa \partial_a \left( \xi^a \psi^i \bar\psi^j \gamma_A \psi^k \bar\psi^l \gamma_B \psi^m 
\cdots \bar\psi^n \gamma_C \psi^p \vert w \vert \right) \right]. 
\label{variation2}
\ea
by using the variations of $\vert w \vert$, i.e. $\delta_\zeta \vert w \vert = \partial_a (\xi^a \vert w \vert)$. 
The variations (\ref{variation1}) and (\ref{variation2}) indicate that the bosonic and fermionic functionals 
in Eqs.(\ref{bosonic}) and (\ref{fermionic}) are linearly exchanged with each other through those variations. 
In fact, the functionals (\ref{bosonic}) and (\ref{fermionic}) lead to LSUSY transformations of component fields 
in (massless) vector linear supermultiplets with general auxiliary fields, 
e.g. in the case for $N = 2$ SUSY in two-dimensional spacetime \cite{ST3,ST4}. 
They are also proper functionals in order to study the NL/LSUSY relations for vector supermultiplets 
in extended SUSY because of the vector components in the functionals $b^i{}_A{}^j$ for $N \ge 2$ SUSY. 

From now on, studying the variations (\ref{variation1}) and (\ref{variation2}) 
by starting with a bosonic component which is defined as $\displaystyle{D = b}$ 
for the lowest-order functional in Eq.(\ref{bosonic}) 
and by examining the commutation relation (\ref{NLSUSYcomm2}) on the functionals (\ref{bosonic}) and (\ref{fermionic}), 
we show that LSUSY transformations of basic bosonic and fermionic components 
defined from those functionals are uniquely determined, 
which satisfy the same commutator algebra as Eq.(\ref{NLSUSYcomm}). 

First, the variation of $D$ is 
\be
\delta_\zeta D 
= -i \bar\zeta^i \!\!\not\!\partial \lambda^i, 
\label{v-D}
\ee
where spinor components $\lambda^i$ are defined as $\lambda^i = f^i$. 
Then, the variations of $\lambda^i$ become 
\be
\delta_\zeta \lambda^i 
= D \zeta^i - {i \over {4 \alpha_{1A}}} \varepsilon_{M1} \gamma^A \!\!\not\!\partial M^i{}_A{}^j \zeta^j, 
\label{v-lambda}
\ee
by using a Fierz transformation. 
In the variation (\ref{v-lambda}), we introduce bosonic components 
$M^i{}_A{}^j = \alpha_{1A} b^i{}_A{}^j$ 
with constants $\alpha_{1A}$, which mean the following kinds of components, 
\ba
\A \A 
M^{ij} = \alpha_{10} b^{ij} = \alpha_{10} \kappa \bar\psi^i \psi^j \vert w \vert, 
\ \ M^i{}_5{}^j = \alpha_{11} b^i{}_5{}^j = i \alpha_{11} \kappa \bar\psi^i \gamma_5 \psi^j \vert w \vert, 
\nonu
\A \A 
M^i{}_a{}^j = \alpha_{12} b^i{}_a{}^j = i \alpha_{12} \kappa \bar\psi^i \gamma_a \psi^j \vert w \vert, 
\ \ M^i{}_{5a}{}^j = \alpha_{13} b^i{}_{5a}{}^j = \alpha_{13} \kappa \bar\psi^i \gamma_5 \gamma_a \psi^j \vert w \vert, 
\nonu
\A \A 
M^i{}_{ab}{}^j = \alpha_{14} b^i{}_{ab}{}^j = \sqrt{2} i \alpha_{14} \kappa \bar\psi^i \sigma_{ab} \psi^j \vert w \vert, 
\label{M-def}
\ea
where $M^i{}_a{}^j$ are the vector components (for $N \ge 2$ SUSY). 
Note that values of the constants $\alpha_{1A}$ in the components (\ref{M-def}) should be determined 
by considering the invariance of actions for vector supermultiplets under LSUSY transformations. 
In addition, in Eq.(\ref{v-lambda}), we use 
$\gamma^A = {\bf 1}, -i \gamma_5, -i \gamma^a, -\gamma_5 \gamma^a \ {\rm or} \ -\sqrt{2} i \sigma^{ab}$ 
and a sign factor $\varepsilon_{M1}$ which appear from the relation 
$\bar\psi^j \gamma_A \psi^i = \varepsilon \bar\psi^i \gamma_A \psi^j$. 
\footnote
{The sign factor $\varepsilon$ is $\varepsilon = +1$ for $\gamma_A = {\bf 1}, i \gamma_5, \gamma_5 \gamma_a$ 
and $\varepsilon = -1$ for $\gamma_A = i \gamma_a, \sqrt{2} i \sigma_{ab}$. 
}
The constants $\alpha$ and the sign factor $\varepsilon$ are also used in the same meanings below. 

Up to the variations of $D$ and $\lambda^i$, their LSUSY transformations 
are unambiguously determined as Eqs.(\ref{v-D}) and (\ref{v-lambda}), 
in which the LSUSY transformation (\ref{v-D}) of $D$ satisfies the commutator algebra (\ref{NLSUSYcomm}) 
under Eq.(\ref{v-lambda}) 
(because of the commutation relation (\ref{NLSUSYcomm2}) on $D = D(\psi^i)$). 

Next, we consider the variations of $M^i{}_A{}^j$ which are calculated as 
\be
\delta_\zeta M^i{}_A{}^j = \alpha_{1A} \left\{ \bar\zeta^i \gamma_A \lambda^j + \bar\lambda^i \gamma_A \zeta^j 
- i \kappa^2 \partial_a \left( \bar\zeta^k \gamma^a \psi^k \bar\psi^i \gamma_A \psi^j \vert w \vert \right) \right\}. 
\label{v-M}
\ee
In the variations (\ref{v-M}), in order to define LSUSY transformations of $M^i{}_A{}^j$ 
by regarding the functionals $f^{ij}{}_A{}^k$ as fermionic components, 
we have to consider deformations of the functionals $\psi^k \bar\psi^i \gamma_A \psi^j$ in the last terms of Eq.(\ref{v-M}). 
Therefore, we examine the commutation relation (\ref{NLSUSYcomm2}) on $\lambda^i$ 
by focusing on two supertranformations of $\lambda^i$ which are obtained 
by means of Eqs.(\ref{v-D}), (\ref{v-lambda}) and (\ref{v-M}) as follows; 
\ba
\delta_{\zeta_1} \delta_{\zeta_2} \lambda^i 
\A = \A \delta_{\zeta_1} D \ \zeta_2^i 
- {i \over {4 \alpha_{1A}}} \varepsilon_{M1} \gamma^A \!\!\not\!\partial \ \delta_{\zeta_1} M^i{}_A{}^j \ \zeta_2^j 
\nonu
\A = \A i \bar\zeta_1^j \gamma^a \zeta_2^j \partial_a \lambda^i 
+ \left[ (1 \leftrightarrow 2)\ {\rm symmetric\ terms\ of}\ \partial_a \lambda \right] 
\nonu
\A \A 
+ {1 \over 16} \varepsilon_{M1} \kappa^2 \gamma^A \!\!\not\!\partial \partial_a 
\left( \bar\zeta_1^k \gamma_B \zeta_2^j \gamma^B \gamma^a \psi^k \bar\psi^i \gamma_A \psi^j \vert w \vert \right). 
\label{twosuper-lambda}
\ea
Since these two supertransformations 
satisfy Eq.(\ref{NLSUSYcomm2}), 
the last terms of Eq.(\ref{twosuper-lambda}) which vanish in the commutation relations 
have to be symmetric under exchanging the indices $1$ and $2$ 
of the spinor transformation parameters ($\zeta_1^k$, $\zeta_2^j$). 
This means that the vanishments of the last terms of Eq.(\ref{twosuper-lambda}) in Eq.(\ref{NLSUSYcomm2}) 
can be confirmed straightforwardly when the $\psi^k$ and $\psi^j$ go into bilinear forms 
$\bar\psi^j \gamma_A \psi^k$ in the last terms of Eqs.(\ref{v-M}) and (\ref{twosuper-lambda}), 
which have the same indices as the spinor transformation parameters 
and reflect the symmetries of the $\bar\zeta_1^k \gamma_B \zeta_2^j$. 

Thus the LSUSY transformations of $M^i{}_A{}^j$ are uniquely determined as 
\be
\delta_\zeta M^i{}_A{}^j 
= \alpha_{1A} \left( \bar\zeta^i \gamma_A \lambda^j + \bar\lambda^i \gamma_A \zeta^j 
+ {i \over {4 \alpha_{2B}}} \varepsilon_{\Lambda1} 
\bar\zeta^k \!\!\not\!\partial \gamma^B \gamma_A \Lambda^{ij}{}_B{}^k \right) 
\label{v-M2}
\ee
by using a Fierz transformation in the last terms of Eq.(\ref{v-M}), 
where we define fermionic components $\Lambda^{ij}{}_A{}^k = \alpha_{2A} f^{ij}{}_A{}^k$ 
with constants $\alpha_{2A}$. 
The LSUSY transformations (\ref{v-lambda}) of $\lambda^i$ satisfy the commutator algebra (\ref{NLSUSYcomm}) 
under Eqs.(\ref{v-D}) and (\ref{v-M2}). 

In accordance with the same argument on the definition of the LSUSY transformations (\ref{v-M2}), 
those of the fermionic components $\Lambda^{ij}{}_A{}^k$ are also derived 
by further introducing bosonic components $C^i{}_A{}^{jk}{}_B{}^l = \alpha_{3AB} b^i{}_A{}^{jk}{}_B{}^l$ 
with constants $\alpha_{3AB}$: Indeed, we consider the variations of $\Lambda^{ij}{}_A{}^k$, 
\ba
\delta_\zeta \Lambda^{ij}{}_A{}^k 
\A = \A \alpha_{2A} \left\{ {1 \over \alpha_{1A}} M^j{}_A{}^k \zeta^i 
- {1 \over {4 \alpha_{1B}}} \gamma^B 
\left( \varepsilon_{M2} \gamma_A \varepsilon'_{M2} M^i{}_B{}^k \zeta^j + \gamma_A \varepsilon'_{M2} M^i{}_B{}^j \zeta^k \right) \right. 
\nonu
\A \A 
\left. - {i \over 4} \kappa^3 \partial_a 
\left( \gamma^B \gamma^a \zeta^l \bar\psi^l \gamma_B \psi^i \bar\psi^j \gamma_A \psi^k \vert w \vert \right) \right\}, 
\label{v-Lambda}
\ea
and deformations of the functionals $\bar\psi^l \gamma_B \psi^i \bar\psi^j \gamma_A \psi^k$ in the last terms of Eq.(\ref{v-Lambda}), 
by examining two supertransformations of $M^i{}_A{}^j$ which satisfy the commutation relation (\ref{NLSUSYcomm2}) 
and which are obtained from Eqs.(\ref{v-lambda}), (\ref{v-M2}) and (\ref{v-Lambda}) as 
\ba
\delta_{\zeta_1} \delta_{\zeta_2} M^i{}_A{}^j 
\A = \A i \bar\zeta_1^k \gamma^a \zeta_2^k \partial_a M^i{}_A{}^j 
+ \left[ (1 \leftrightarrow 2)\ {\rm symmetric\ terms\ of}\ D \ {\rm and}\ \partial_a M^i{}_A{}^j \right] 
\nonu
\A \A 
+ {1 \over 16} \varepsilon_{\Lambda1} \kappa^3 \partial_a \partial_b 
\left( \bar\zeta_2^k \gamma^a \gamma^B \gamma_A 
\gamma^C \gamma^b \zeta_1^l \bar\psi^l \gamma_C \psi^i \bar\psi^j \gamma_B \psi^k \vert w \vert \right). 
\label{twosuper-M}
\ea
Because of the indices of the spinor transformation parameters ($\zeta_1^l$, $\zeta_2^k$) 
in the last terms of Eq.(\ref{twosuper-M}) 
which vanish in the commutation relation, the $\psi^l$ and $\psi^k$ have to take bilinear forms $\bar\psi^k \gamma_A \psi^l$ 
in the last terms of Eqs.(\ref{v-Lambda}) and (\ref{twosuper-M}) 
in order to confirm straightforwardly the vanishments of the last terms of Eq.(\ref{twosuper-M}) in Eq.(\ref{NLSUSYcomm2}). 

Therefore, the LSUSY transformations of $\Lambda^{ij}{}_A{}^k$ 
are given 
as 
\ba
\delta_\zeta \Lambda^{ij}{}_A{}^k 
\A = \A \alpha_{2A} \left\{ {1 \over \alpha_{1A}} M^j{}_A{}^k \zeta^i 
- {1 \over {4 \alpha_{1B}}} \gamma^B \left( \varepsilon_{M2} \gamma_A \varepsilon'_{M2} M^i{}_B{}^k \zeta^j 
+ \gamma_A \varepsilon'_{M2} M^i{}_B{}^j \zeta^k \right) \right. 
\nonu
\A \A 
\left. + {i \over {16 \alpha_{3ACB}}} \varepsilon_{C1} \varepsilon'_{C1} 
\gamma^B \!\!\not\!\partial C^{iCjk}{}_{ACB}{}^l \zeta^l \right\}, 
\label{v-Lambda2}
\ea
by using a Fierz transformation in the last terms of Eq.(\ref{v-Lambda}). 
In Eq.(\ref{v-Lambda2}) we define $C^{iCjk}{}_{ACB}{}^l = \alpha_{3ACB} b^{iCjk}{}_{ACB}{}^l$ 
\ $(= \alpha_{3ACB} \kappa^3 \bar\psi^i \gamma^C \psi^j \bar\psi^k \gamma_A \gamma_C \gamma_B \psi^l \vert w \vert)$ 
with constants $\alpha_{3ACB}$ for convenience, which can be expanded by means of the components $C^i{}_A{}^{jk}{}_B{}^l$ 
under the Clifford algebra for $\gamma$ matrices. 
Then, the LSUSY transformations (\ref{v-M2}) of $M^i{}_A{}^j$ 
satisfy the commutator algebra (\ref{NLSUSYcomm}) under Eqs.(\ref{v-lambda}) and (\ref{v-Lambda2}). 

As for the variations of $C^i{}_A{}^{jk}{}_B{}^l$ 
and the higher-order functionals of $\psi^i$ in Eqs.(\ref{bosonic}) and (\ref{fermionic}), 
two additional problems appear in the definition of their LSUSY transformations. 
In the case of $C^i{}_A{}^{jk}{}_B{}^l$, let us explicitly show that those problems are also solved 
by examining two supertransformations of the components $\Lambda^{ij}{}_A{}^k$ and $C^i{}_A{}^{jk}{}_B{}^l$ 
based on the commutation relation (\ref{NLSUSYcomm2}). 
In the variations of $C^i{}_A{}^{jk}{}_B{}^l$, 
\ba
\delta_\zeta C^i{}_A{}^{jk}{}_B{}^l 
\A = \A \alpha_{3AB} \kappa^2 \left[ \left\{ 
\left( \bar\zeta^i \gamma_A \psi^j + \bar\psi^i \gamma_A \zeta^j \right) \bar\psi^k \gamma_B \psi^l 
\right. \right. 
\nonu
\A \A 
\hspace{1.5cm}
\left. \left. 
+ \bar\psi^i \gamma_A \psi^j \left( \bar\zeta^k \gamma_B \psi^l + \bar\psi^k \gamma_B \zeta^l \right) 
\right\} \vert w \vert \right. 
\nonu
\A \A 
\left. - i \kappa^2 \partial_a 
\left( \bar\zeta^m \gamma^a \psi^m \bar\psi^i \gamma_A \psi^j \bar\psi^k \gamma_B \psi^l \vert w \vert \right) \right], 
\label{v-C}
\ea
one problem is how LSUSY transformations of $C^i{}_A{}^{jk}{}_B{}^l$ are determined 
with respect to the fermionic components $\Lambda^{ij}{}_A{}^k$. 
It is solved by focusing on derivative terms of $\Lambda^{ij}{}_A{}^k$ ($\partial \Lambda$-terms) 
in two supertransformations of $\Lambda^{ij}{}_A{}^k$, 
the parts of which are obtained from the LSUSY transformations (\ref{v-M2}) and (\ref{v-Lambda2}); 
namely, those $\partial \Lambda$-terms are 
\ba
\A \A 
\delta_{\zeta_1} \delta_{\zeta_2} \Lambda^{ij}{}_A{}^k 
\left[ \partial \Lambda \ {\rm terms\ obtained\ through}\ \delta_\zeta M \right] 
\nonu
\A \A 
\hspace{7mm} 
= {i \over 4} \alpha_{2A} \left[ {1 \over \alpha_{2B}} \varepsilon_{\Lambda1} 
\partial_a \left( \bar\zeta_1^l \gamma^a \gamma^B \gamma_A \Lambda^{jk}{}_B{}^l \zeta_2^i \right) \right. 
\nonu
\A \A 
\hspace{1cm}
- {1 \over {4 \alpha_{2C}}} \varepsilon'_{M2} \varepsilon_{\Lambda1} \gamma^B 
\left\{ \varepsilon_{M2} \gamma_A \partial_a \left( \bar\zeta_1^l \gamma^a \gamma^C \gamma_B \Lambda^{ik}{}_C{}^l \zeta_2^j \right) 
\right. 
\nonu
\A \A 
\hspace{4.5cm}
\left. + \gamma_A \partial_a \left( \bar\zeta_1^l \gamma^a \gamma^C \gamma_B \Lambda^{ij}{}_C{}^l \zeta_2^k \right) 
\right\} \bigg], 
\label{twosuper-Lambda}
\ea
where the $\partial_a (\Lambda^{jk}{}_A{}^l, \Lambda^{ik}{}_A{}^l, \Lambda^{ij}{}_A{}^l)$-type terms appear. 
Based on the commutation relation (\ref{NLSUSYcomm2}) on $\Lambda^{ij}{}_A{}^k$, 
the $\partial \Lambda$-terms of Eq.(\ref{twosuper-Lambda}) 
have to cancel with ones which are obtained from LSUSY transformations of $C^i{}_A{}^{jk}{}_B{}^l$ in the commutation relation. 
Therefore, the variations (\ref{v-C}) have to give $\Lambda$-terms with the same arrangement of the internal indices 
as in Eq.(\ref{twosuper-Lambda}) in order to realize straightforwardly the cancellations of the $\partial \Lambda$-terms.  

Thus the LSUSY transformations of $C^i{}_A{}^{jk}{}_B{}^l$ are defined 
with respect to the fermionic components $\Lambda^{ij}{}_A{}^k$ as 
\ba
\delta_\zeta C^i{}_A{}^{jk}{}_B{}^l \left[ \Lambda \ {\rm terms} \right] 
\A = \A \alpha_{3AB} \left\{ {1 \over \alpha_{2B}} \left( \bar\zeta^i \gamma_A \Lambda^{jk}{}_B{}^l 
+ \varepsilon_{\Lambda2} \bar\zeta^j \gamma_A \Lambda^{ik}{}_B{}^l \right) \right. 
\nonu
\A \A 
\left. - {1 \over {4 \alpha_{2C}}} \varepsilon_{\Lambda2} 
\left( \bar\zeta^k \gamma_B \gamma^C \gamma_A \Lambda^{ij}{}_C{}^l 
+ \varepsilon'_{\Lambda2} \bar\zeta^l \gamma_B \gamma^C \gamma_A \Lambda^{ij}{}_C{}^k \right) \right\}, 
\label{v-C2}
\ea
by using Fierz transformations in the variations (\ref{v-C}). 
Note that the last terms for $\Lambda^{ij}{}_C{}^k$ in Eq.(\ref{v-C2}) 
give the translations of $\Lambda^{ij}{}_A{}^k$ (i.e., $\Xi^a \partial_a \Lambda^{ij}{}_A{}^k$) 
in a commutator algebra for the LSUSY transformations (\ref{v-Lambda2}). 

Another problem in the variation (\ref{v-C}) is how LSUSY transformations of $C^i{}_A{}^{jk}{}_B{}^l$ are determined 
with respect to new fermionic components defined from the functionals $f^{ij}{}_A{}^{kl}{}_B{}^m$. 
Following the argument on the definition of the LSUSY transformations (\ref{v-M2}) and (\ref{v-Lambda2}), 
we can rewrite the last terms for $f^{mi}{}_A{}^{jk}{}_B{}^l$ in the variations (\ref{v-C}) 
by means of a Fierz transformation as follows; 
\be
\delta_\zeta C^i{}_A{}^{jk}{}_B{}^l \left[ \partial_a f^{ki}{}_A{}^{jl}{}_B{}^m \ {\rm terms} \right] 
= {i \over 4} \varepsilon_{\Psi1} \alpha_{3AB} \kappa^4 \partial_a 
\left( \bar\zeta^m \gamma^a \gamma_C \gamma_B \psi^k \bar\psi^i \gamma_A \psi^j 
\bar\psi^l \gamma^C \psi^m \vert w \vert \right). 
\label{v-C3}
\ee
By examining the two supertransformations of $\Lambda^{ij}{}_A{}^k$ 
with respect to second-order derivative terms of the functionals $f^{ki}{}_A{}^{jl}{}_B{}^m$ of Eq.(\ref{v-C3}), 
those terms vanish in the commutation relation (\ref{NLSUSYcomm2}) on $\Lambda^{ij}{}_A{}^k$ 
thanks to the symmetries of the indices $m$ and $l$ in the functionals, 
which correspond to the indices of the spinor transformation parameters $(\zeta_1^m, \zeta_2^l)$ 
in the two supertransformations. 

However, in order to define LSUSY transformations of $C^i{}_A{}^{jk}{}_B{}^l$, 
considering deformation of $\psi^k \bar\psi^i \gamma_A \psi^j$ in the functionals $f^{ki}{}_A{}^{jl}{}_B{}^m$ 
of Eq.(\ref{v-C3}) is a remaining problem. 
To solve this, two supertransformations of $C^i{}_A{}^{jk}{}_B{}^l$ have to be studied 
together with the definition of LSUSY transformations for the functionals $f^{ij}{}_A{}^{kl}{}_B{}^m$ 
with respect to $C^i{}_A{}^{jk}{}_B{}^l$: 
Indeed, in two supertransformations of $C^i{}_A{}^{jk}{}_B{}^l$, 
derivative terms of $C^i{}_A{}^{jk}{}_B{}^l$ ($\partial C$-terms) 
which are obtained from the LSUSY transformations (\ref{v-Lambda2}) and (\ref{v-C2}) are 
\ba
\A \A 
\delta_{\zeta_1} \delta_{\zeta_2} C^i{}_A{}^{jk}{}_B{}^l 
\left[ \partial C \ {\rm terms\ obtained\ through}\ \delta_\zeta \Lambda \right] 
\nonu
\A \A 
\hspace{7mm} 
= {{i \alpha_{3AB}} \over {16 \alpha_{3ABC}}} \varepsilon_{C1} \varepsilon'_{C1} \left\{ 
\left( \bar\zeta_2^i \gamma_A \gamma^C \!\!\not\!\partial C^{jDkl}{}_{BDC}{}^m \zeta_1^m 
+ \varepsilon_{\Lambda2} \bar\zeta_2^j \gamma_A \gamma^C \!\!\not\!\partial C^{iDkl}{}_{BDC}{}^m \zeta_1^m \right) 
\right. 
\nonu
\A \A 
\hspace{3.5cm} 
- {1 \over 4} \varepsilon_{\Lambda2} 
\left( \bar\zeta_2^k \gamma_B \gamma_C \gamma_A \gamma^D \!\!\not\!\partial C^{iEjlC}{}_{ED}{}^m \zeta_1^m \right. 
\nonu
\A \A 
\hspace{5cm}
\left. + \varepsilon'_{\Lambda2} \bar\zeta_2^l \gamma_B \gamma_C \gamma_A \gamma^D 
\!\!\not\!\partial C^{iEjkC}{}_{ED}{}^m \zeta_1^m \right) \bigg\}, 
\label{twosuper-C}
\ea
where the $\partial_a (C^j{}_A{}^{kl}{}_B{}^m, C^i{}_A{}^{kl}{}_B{}^m, 
C^i{}_A{}^{jl}{}_B{}^m, C^i{}_A{}^{jk}{}_B{}^m)$-type terms appear. 

In accordance with the internal indices of $\partial C$-terms in Eq.(\ref{twosuper-C}), 
if we define fermionic components $\Psi^{ij}{}_A{}^{kl}{}_B{}^m = \alpha_{4AB} f^{ij}{}_A{}^{kl}{}_B{}^m$ 
with constants $\alpha_{4AB}$, LSUSY transformations of $\Psi^{ij}{}_A{}^{kl}{}_B{}^m$ 
have to be determined with respect to $C^i{}_A{}^{jk}{}_B{}^l$ 
as 
\ba
\A \A 
\delta_\zeta \Psi^{ij}{}_A{}^{kl}{}_B{}^m \left[ C \ {\rm terms} \right] 
\nonu
\A \A 
= \alpha_{4AB} \left\{ {1 \over \alpha_{3AB}} \zeta^i C^j{}_A{}^{kl}{}_B{}^m \right. 
\nonu
\A \A 
\hspace{3mm} - {1 \over {4 \alpha_{3CB}}} 
\left( \varepsilon_{C2} \gamma_C \gamma_A \zeta^j \varepsilon'_{C2} C^{iCkl}{}_B{}^m 
+ \gamma_C \gamma_A \zeta^k \varepsilon'_{C2} C^{iCjl}{}_B{}^m \right) 
\nonu
\A \A 
\hspace{3mm} \left. + {1 \over {16 \alpha_{3ADC}}} 
\left( \varepsilon_{C3} \gamma_C \gamma_B \zeta^l \varepsilon'_{C3} \varepsilon''_{C3} C^i{}_D{}^{jk}{}_A{}^{DCm} 
+ \gamma_C \gamma_B \zeta^m \varepsilon'_{C3} \varepsilon''_{C3} C^i{}_D{}^{jk}{}_A{}^{DCl} \right) \right\}, 
\label{v-Psi}
\ea
which are given by using Fierz transformations in the variations of $\Psi^{ij}{}_A{}^{kl}{}_B{}^m$. 
Note that $\Lambda$-terms in the commutation relation (\ref{NLSUSYcomm2}) on $\Psi^{ij}{}_A{}^{kl}{}_B{}^m$ vanish 
only if the LSUSY transformations (\ref{v-Psi}) are defined, 
since the LSUSY ones of $C^i{}_A{}^{jk}{}_B{}^l$ have already been determined as Eq.(\ref{v-C2}). 

Then, in the commutation relation (\ref{NLSUSYcomm2}) on $C^i{}_A{}^{jk}{}_B{}^l$, 
the $\partial C$-terms of Eq.(\ref{twosuper-C}) cancel with those terms 
obtained through the LSUSY transformations (\ref{v-Psi}), 
when LSUSY transformations of $C^i{}_A{}^{jk}{}_B{}^l$ 
are defined with respect to $\Psi^{ij}{}_A{}^{kl}{}_B{}^m$ as 
\be
\delta_\zeta C^i{}_A{}^{jk}{}_B{}^l \left[ \Psi \ {\rm terms} \right] 
= - {{i \alpha_{3AB}} \over {16 \alpha_{4AB}}} \varepsilon_{\Psi1} \varepsilon_{\Psi2} 
\bar\zeta^m \!\!\not\!\partial \gamma^C \gamma_B \gamma^D \gamma_A \Psi^{ij}{}_D{}^{kl}{}_C{}^m. 
\label{v-C4}
\ee
Here we also note that the last terms with respect to $C^i{}_D{}^{jk}{}_A{}^{DCl}$ 
in the LSUSY transformations (\ref{v-Psi}) give the translations of $C^i{}_A{}^{jk}{}_B{}^l$ 
(i.e., $\Xi^a \partial_a C^i{}_A{}^{jk}{}_B{}^l$) 
in a commutator algebra for the LSUSY transformations of $C^i{}_A{}^{jk}{}_B{}^l$ through Eq.(\ref{v-C4}). 

Thus the LSUSY transformations of $C^i{}_A{}^{jk}{}_B{}^l$ 
are uniquely determined as Eqs.(\ref{v-C2}) and (\ref{v-C4}). 
The LSUSY transformations (\ref{v-Lambda2}) of $\Lambda^{ij}{}_A{}^k$ 
satisfy the commutator algebra (\ref{NLSUSYcomm}) under Eqs.(\ref{v-M2}), (\ref{v-C2}) and (\ref{v-C4}). 
As for LSUSY transformations of components for higher-order functionals of $\psi^i$ 
than $b^i{}_A{}^{jk}{}_B{}^l$ ($C^i{}_A{}^{jk}{}_B{}^l$), 
they can be determined in accordance with the above arguments for the definition of LSUSY transformations, 
which terminates with those of bosonic components for the functionals at ${\cal O}\{ (\psi^i)^{4N} \}$ in Eq.(\ref{bosonic}). 

We summarize our results as follows. 
In this letter, we have discussed a linearization procedure of NLSUSY 
based on the commutatation relation (\ref{NLSUSYcomm2}) on the functionals (\ref{functionals}) 
in terms of the NG fermions $\psi^i$ and their derivative terms under the NLSUSY transformations (\ref{NLSUSY}). 
In the case of the bosonic and fermionic functionals (\ref{bosonic}) and (\ref{fermionic}), 
we have shown that the general LSUSY transformations of the basic components defined from those functionals 
are uniquely determined in the linearization procedure; 
indeed, the variations (\ref{variation1}) and (\ref{variation2}) have been studied 
by starting with the scalar component $D$ for the lowest-order bosonic functional of $\psi^i$. 
The LSUSY transformations of $D$ and the spinor components $\lambda^i$ 
are unambiguously determined as Eqs.(\ref{v-D}) and (\ref{v-lambda}). 
As for the components $M^i{}_A{}^j$ and $\Lambda^{ij}{}_A{}^k$, their LSUSY transformations 
are uniquely determined, in particular, with respect to the $(\partial \Lambda, \partial C)$-terms 
in Eqs.(\ref{v-M2}) and (\ref{v-Lambda2}) by examining the $(\partial^2 \Lambda, \partial^2 C)$-terms 
in the two supertransformtions (\ref{twosuper-lambda}) and (\ref{twosuper-M}). 

We have also explained the derivation of the LSUSY transformations of the components $C^i{}_A{}^{jk}{}_B{}^l$ 
as a general case in the linearization procedure: 
The LSUSY transformations (\ref{v-C2}) to $\Lambda$-terms are obtained 
by considering the $\partial \Lambda$-terms in the two supertransformations (\ref{twosuper-Lambda}). 
On the other hand, the LSUSY transformations to $\partial \Psi$-terms 
have to be determined not as Eq.(\ref{v-C3}) which is obtained by considering the $\partial^2 \Psi$-terms 
in the two supertransformtions of $\Lambda^{ij}{}_A{}^k$ 
but as Eq.(\ref{v-C4}) which is defined by further examining the variations of $\Psi^{ij}{}_A{}^{kl}{}_B{}^m$ 
with respect to the $C$-terms through the two supertransformations (\ref{twosuper-C}). 

These results for the basic components which are defined straightforwardly from the functionals 
(\ref{bosonic}) and (\ref{fermionic}) show that their LSUSY transformations are uniquely determined 
in the linearization of NLSUSY by examining the two supertransformations of them, 
i.e. the commutation relation (\ref{NLSUSYcomm2}) from on the lowest-order functional 
to on the same-order ones, including up to the vanishing terms in the commutation relation. 

Finally, we mention transitions from the basic components in the general LSUSY transformations 
to component fields in (massless) vector supermultiplets prior to transforming to gauge supermultiplets. 
For $N = 1$ SUSY, the bosonic and fermionic functionals (\ref{bosonic}) and (\ref{fermionic}) reduce to 
\footnote
{
In the reduction of the functionals (\ref{bosonic}) and (\ref{fermionic}) to Eq.(\ref{N1-functional}), 
we have used identities which are obtained by using Fierz transformations 
as $\psi \bar\psi \gamma_5 \gamma_a \psi = - \gamma_5 \gamma_a \psi \bar\psi \psi$, 
$\gamma_5 \psi \bar\psi \gamma_5 \psi = - \psi \bar\psi \psi$, 
$\bar\psi \gamma_5 \gamma_a \psi \bar\psi \gamma_5 \gamma_b \psi = \eta_{ab} \bar\psi \psi \bar\psi \psi$ 
and $\bar\psi \gamma_5 \psi \bar\psi \gamma_5 \psi = - \bar\psi \psi \bar\psi \psi$, etc. 
}
\ba
\A \A 
b = \kappa^{-1} \vert w \vert, 
\ \ f = \psi \vert w \vert, 
\ \ b_1 = \kappa \bar\psi \psi \vert w \vert, 
\ \ b_5 = i \kappa \bar\psi \gamma_5 \psi \vert w \vert, 
\nonu
\A \A 
b_{5a} = \kappa \bar\psi \gamma_5 \gamma_a \psi \vert w \vert, 
\ \ f_1 = \kappa^2 \psi \bar\psi \psi \vert w \vert, 
\ \ b_{11} = \kappa^3 \bar\psi \psi \bar\psi \psi, 
\label{N1-functional}
\ea
and the basic component fields are defined from the set of the functionals (\ref{N1-functional}) as 
\ba
\A \A 
D = b(\psi), 
\ \ \lambda = f(\psi), 
\ \ A = \alpha_1 b_1(\psi), 
\ \ B = \alpha_2 b_5(\psi), 
\nonu
\A \A
v_a = \alpha_3 b_{5a}(\psi), 
\ \ \Lambda = \alpha_4 f_1(\psi), 
\ \ C = \alpha_5 b_{11}(\psi), 
\label{N1-comp}
\ea
with the constants $\alpha_m \ (m = 1,2,\cdots,5)$, though the "vector" field $v_a$ defined in Eq.(\ref{N1-comp}) 
is expressed in terms of the axial-vector functional $b_{5a}$ \cite{STT1}. 
Note that the degrees of freedom of the above bosonic and fermionic components for $N = 1$ SUSY are balanced as $8 = 8$ 
and the basic component fields in $N = 1$ SUSY theories are defined, in general, by further multiplying 
those components by an overall constant $\xi$ which gives a vacuum expectation value of the $D$-term. 

From the functional representation of the basic component fields (\ref{N1-comp}), 
we have recently shown \cite{MT} that LSUSY transformations for a $N = 1$ vector supermultiplet 
with the general auxiliary fields $(\Lambda, C)$ \cite{WZ,WB} are derived 
by using the commutator-based linearization procedure in this letter. 
In addition, both $U(1)$ gauge and scalar supermultiplets in $N = 1$ SUSY theories are also constructed 
from the same set of the functionals (\ref{N1-functional}) by means of appropriate recombinations 
of the basic components (\ref{N1-comp}) as follows \cite{MT}; namely, recombinations, 
\be
\tilde D = \left( D + {1 \over {8 \alpha_5}} \Box C \right)(\psi), 
\ \ \tilde \lambda = \left( \lambda + {i \over {2 \alpha_4}} \!\!\not\!\partial \Lambda \right)(\psi), 
\label{recombi-gauge}
\ee
lead to the $U(1)$ gauge supermultiplet with the component fields $(\tilde D, \tilde \lambda, v_a)$, 
whereas ones, 
\be
F = \left( D - {1 \over {8 \alpha_5}} \Box C \right)(\psi), 
\ \ G = \partial^a v_a(\psi), 
\ \ \chi = \left( \lambda - {i \over {2 \alpha_4}} \!\!\not\!\partial \Lambda \right)(\psi), 
\label{recombi-scalar}
\ee
give the scalar supermultiplet with the component fields $(F, G, \chi, A, B)$. 
Here we also note that values of the constants $\alpha_m$ in the components (\ref{N1-comp}) 
are given from the LSUSY invariance (and the definition) of (free) actions 
as $\displaystyle{\alpha_3^2 = {1 \over 4}}$ \Big( $\displaystyle{\alpha_4 = {1 \over 2}}$, 
\ $\displaystyle{\alpha_5 = {1 \over 8}}$ \Big) 
for the $U(1)$ gauge supermultiplet and as $\displaystyle{\alpha_1^2 = \alpha_2^2 = \alpha_3^2 = {1 \over 4}}$ 
\Big( $\displaystyle{\alpha_4 = -{1 \over 2}}$, \ $\displaystyle{\alpha_5 = -{1 \over 8}}$ \Big) for the scalar one. 

Moreover, in $N = 1$ SUSY theories, a constrained-superfield NLSUSY action of a goldstino 
for low-energy effective theories was constructed in Ref.\cite{KoS}, 
in which a spinor field $g$ identified as the goldstino and the (auxiliary) scalar $F$-term constitute 
a nonlinear supermultiplet through a chiral superfield $X$ with a constraint $X^2 = 0$. 
In the constrained chiral superfield, a scalar component (i.e., a superpartner of the goldstino) 
is represented in terms of $g$ (which corresponds to $\chi = \chi(\psi)$) 
and the auxiliary scalar field $F_c \ (= F(\psi) + iG(\psi))$ as a composite $\displaystyle{g^2 \over {2F_c}}$ 
(in the two-component spinor notation). 
From the viewpoint of the NG-fermion functional expression (\ref{recombi-scalar}) for the scalar supermultiplet, 
the composite scalar field in $X$ with $X^2 = 0$ relates to a complex scalar field $\phi$ 
defined from the scalar components $(A,B)$ in Eq.(\ref{N1-comp}); 
its relation is confirmed easily at least at a leading order for the composite scalar as 
\be
{g^2 \over {2F_c}}(\psi) \Bigg\vert_{\rm leading} 
= { \{ (\chi) \vert_\lambda \}^2 \over {2(F) \vert_D}}(\psi) = {\lambda^2 \over {2 D}}(\psi) 
= {1 \over 2} \kappa \psi^2 \vert w \vert = (A+iB)(\psi) = \phi(\psi). 
\label{KS-scalar}
\ee
The all-order NG-fermion functional correspondance for Eq.(\ref{KS-scalar}) is also expected 
from the equivalence of the VA and Komargodski-Seiberg NLSUSY actions, which has been shown explicitly in Ref.\cite{Zh}, 
in addition to the NL/LSUSY relation for the $N = 1$ scalar supermultiplet. 

For $N = 2$ SUSY, we expect that component fields in vector and gauge supermultiplets 
are defined by multiplying the basic components by overall constants $\xi^I$ ($I = 1,2,3$) which imply a $SU(2)$-structure 
in functionals (composites) of the NG fermions for chiral fermions and in LSUSY transformations of $D$-terms \cite{STT2}, 
but their explicit derivations are under investigation. 
The invariance of fundamental actions in LSUSY theories for $N \ge 2$ SUSY 
would be explicitly confirmed after transitions to vector and gauge supermultiplets 
which depend on both the basic components and overall constants with internal indices. 

\vspace{5mm}

\noindent
{\large\bf Acknowledgements} \\[2mm]
I am very grateful to Kazunari Shima for useful and stimulating discussions.

\newpage

%
\newcommand{\NP}[1]{{\it Nucl.\ Phys.\ }{\bf #1}}
\newcommand{\PL}[1]{{\it Phys.\ Lett.\ }{\bf #1}}
\newcommand{\CMP}[1]{{\it Commun.\ Math.\ Phys.\ }{\bf #1}}
\newcommand{\MPL}[1]{{\it Mod.\ Phys.\ Lett.\ }{\bf #1}}
\newcommand{\IJMP}[1]{{\it Int.\ J. Mod.\ Phys.\ }{\bf #1}}
\newcommand{\PR}[1]{{\it Phys.\ Rev.\ }{\bf #1}}
\newcommand{\PRL}[1]{{\it Phys.\ Rev.\ Lett.\ }{\bf #1}}
\newcommand{\PTP}[1]{{\it Prog.\ Theor.\ Phys.\ }{\bf #1}}
\newcommand{\PTPS}[1]{{\it Prog.\ Theor.\ Phys.\ Suppl.\ }{\bf #1}}
\newcommand{\AP}[1]{{\it Ann.\ Phys.\ }{\bf #1}}

\end{document}